# Improving Hard Disk Contention-based Covert Channel in Cloud Computing Environment

Bartosz Lipiński, Wojciech Mazurczyk, Krzysztof Szczypiorski
Warsaw University of Technology, Institute of Telecommunications
Warsaw, Poland
e-mail: B.Lipinski@stud.elka.pw.edu.pl, {wm, ksz}@tele.pw.edu.pl

*Abstract*—Steganographic methods allow the covert exchange of secret data between parties aware of the procedure. The cloud computing environment is a new and hot target for steganographers, and currently not many solutions have been proposed. This paper proposes CloudSteg which is a steganographic method that allows the creation of a covert channel based on hard disk contention between the two cloud instances that reside on the same physical machine. Experimental results conducted using open source cloud environment OpenStack, show that CloudSteg is able to achieve a bandwidth of about 0.1 bps which is 1000 times higher than is known from the state-of-the-art version.

## I. INTRODUCTION

Cloud computing is a very popular and still evolving paradigm. It allows customers of cloud providers to avoid start-up costs and/or reduce operating costs. It also increases their flexibility by immediately acquiring services and infrastructural resources when needed. NIST (National Institute of Standards and Technology) defines cloud computing as [1]: "a model for enabling ubiquitous, convenient, on-demand network access to a shared pool of configurable computing resources (e.g., networks, servers, storage, applications, and services) that can be rapidly provisioned and released with minimal management effort or service provider interaction". Key characteristics of cloud computing include: on-demand self-service, broad network access, resource pooling, rapid elasticity, and measured service. NIST also defines three service models in which cloud users control:

1. Only user-specific application configurations, known as *SaaS* (Software as a Service),
2. Deployed applications and possibly application hosting environment configurations, known as *PaaS* (Platform as a Service),
3. Everything (e.g. operating systems, storage, deployed applications) except the underlying cloud infrastructure, known as *IaaS* (Infrastructure as a Service).

Steganographic techniques can be used to provide a perfect tool for data exfiltration, to enable network attacks or hidden communication among secret parties. The aim of these techniques is to hide secret data in the innocent looking carrier e.g. in normal transmissions of users. It is important to note that steganography embraces all the *methods* that can be used to create *covert channels* in a networking environment. Depending on the extent of the created covert channel it can be *network* (allowing to communicate secretly through the network) or local (the extent is limited to the single physical machine) [16].

In an ideal situation such hidden data exchange cannot be detected by third parties. The best carrier for secret data must possess two features: it should be popular i.e. usage of such a carrier should not be considered as an anomaly itself, and modification of the carrier related to inserting the steganogram should not be "visible" to a third party not aware of the steganographic procedure.

Every steganographic method can be described typically by the following set of characteristics [16]: its steganographic bandwidth, undetectability, and robustness. The term "steganographic bandwidth" refers to the amount of secret data that can be sent per unit time when using a particular method. Undetectability is defined as the inability to detect a steganogram within a certain carrier. The most popular way to detect a steganogram is to analyse the statistical properties of the captured data and compare them with the typical values for that carrier. The last characteristic, robustness, is defined as the amount of alteration a steganogram can withstand without secret data being destroyed.

Of course a good steganographic method should be as robust and hard to detect as possible while offering the highest bandwidth. However it must be noted that there is always a fundamental trade-off among these three measures that is necessary.

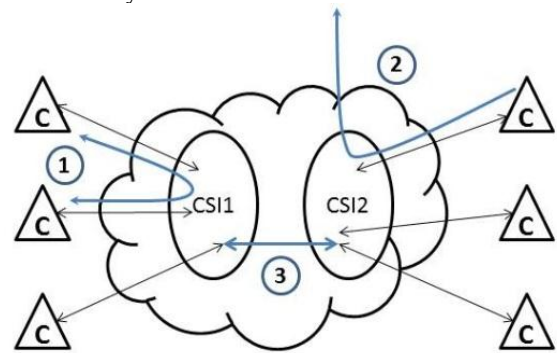

Fig. 1. Hidden communication scenarios for cloud computing environment [13] (CSI – Cloud Service Instance, C – Cloud Clients/Customers)

In the Internet today we witness expansion of different, advanced services from which more and more are migrating to cloud computing services. The major cloud service providers are significantly investing in their infrastructure and in acquiring customers. The important players' list includes: Google (Gmail, GoogleDocs), Microsoft (Azure), Amazon (Amazon Web Services), Cisco (WebEx). These services utilise sometimes complex protocols and infrastructures to achieve their goals. Therefore, they can be considered as perfect candidates for secret data carriers because, generally, the more complex protocols/services are, the more opportunities for information hiding. Indeed the cloud computing environment becomes a new and hot target for steganographers and currently not many information hiding solutions have been proposed ([4-11]).

In [13] we proposed to consider three types of steganographic communication in cloud computing based on where the secret data receiver (SR) is placed (Fig. 1). All these scenarios require, of course, hiding intentions with which cloud services will be used:

(1) SR is situated in the same cloud instance as the secret data sender (SS). This case considers the scenario in which two cloud service users exchange confidential data for which at least one of them is unauthorised. Such communication includes utilisation of the so-called *network steganography* methods that use network protocols (data units or the way they are exchanged), or relations between two or more protocols as a hidden data carrier [14]. In this case a *network covert channel* is created.

(2) SR is situated outside the cloud service environment and cloud service capabilities are used to perform network attacks. In this case steganographic communication can be used e.g. to coordinate cloud botnet or malware [2]. This type of communication may also include utilisation of network steganographic techniques. In this case a *network covert channel* is created.

(3) SR is situated in a different cloud instance than SS but in the same cloud service environment. Realization of such covert communication is typically possible by utilisation of the common shared resources for steganographic purposes and they can relay e.g. on data cross deduplication mechanism [3] or caches shared by services instances on the same physical machines [7]. In this case a *local covert channel* is created.

In this paper we investigate the last scenario. We focus on improving the steganographic methods that allow the creation of local covert channels. The first work that initiated discussion on covert communication in the cloud computing environment was introduced by Ristenpart et al. [7] in 2009. Among many identified covert channels there, in which two cloud instances cooperate to send secret messages via a shared resource, authors describe a method that relies on hard disk contention between co-resident instances. The obtained steganographic bandwidth for this solution in their work was 0.0005 bps (however, it must be noted that authors made no attempts to optimize it, as it was utilized to determine co-residence of the instances).

Therefore the **main contribution** of this paper is to show to what extent the hard disk contention-based method can be further improved. Our experimental results suggest that it is able to achieve a bandwidth of about 0.1 bps which is 1000 times higher than previously reported in [7]. The improvement is achieved mainly by applying three synchronization mechanisms (compare Section IV B):
(1) First, that allows to discover the beginning of the transmitted secret data bit,
(2) Symbol synchronization mechanism,
(3) Frame synchronization mechanism.

The process of exchanging covert data between SS and SR in CloudSteg is fully automated, and proof-of-concept implementation is realized and evaluated in an open source OpenStack cloud environment [12].

The rest of this paper is structured as follows. Section II presents the state-of-the-art in information hiding for cloud computing environment. In Section III threat model and assumptions are pointed out. Section IV describes CloudSteg in details and Section V presents its experimental evaluation. Finally, Section VI concludes our work.

## II. RELATED WORK

Information hiding in a cloud computing environment is quite a new research field which has already attracted the attention of the research community.

As mentioned in the previous section, the first paper that started discussion on potential opportunities for hidden communication in cloud computing was work published in 2009 by Ristenpart et al. [7]. Many of the consecutive papers greatly relied on the foundations that this work provided. Its main aim was not solely focused on identifying possible covert channels. The goal was rather to present means to determine co-residence of the instances in a public Amazon's Elastic Compute Cloud (EC2) service. EC2 allows users to rent its computational resources capabilities and utilize them for their own use. After the co-residence is established cross-virtual machines information leakage is proved to be possible based on the commonly shared resources contention. This is how a local covert channel is created that enables one way (side channel) or both way data exchange.

In the rest of this section we review the existing information hiding methods based on what shared resource is utilized as a hidden data carrier: CPU, RAM or hard disk, consecutively.

Utilization of shared CPU was first proposed by Ristenpart et al. [7]. Authors introduced a steganographic technique based on CPU L2 cache. L2 cache is utilized to store recently accessed information between L1 cache and RAM memory. The key idea of the proposed method to enable secret data exchange between co-resident cloud instances is to access frantically L2 cache when binary '1' is to be sent, and remain idle in case binary '0' is to be transmitted. The experimentally obtained steganographic bandwidth of such method was about 0.2 bps.

Next, Xu et al. [10] proposed improvements of this method by adjusting this technique to new policies in Amazon EC2 and increasing the available steganographic bandwidth to about 3.2 bps on average.

In [6] Okamura et al. designed and evaluated a similar approach which utilized the load of a shared CPU to encode secret data bits. The resulting steganographic bandwidth was about 2 bps. However this work has little practical applicability as it only works under the assumption that both colluding cloud instances share the same processor's physical core, which for modern servers based on multi-core processors is hard to achieve. A similar approach, but utilizing the SMT (Simultaneous Multi-Threading) technique, was proposed by Wang and Lee [8]. The drawback of this method is that it can be only applied to machines that support SMT.

Zhang et al. [11] monitor the patterns of L2 cache usage within a guest domain to build a classifier of the usage patterns. Such an approach allows the utilization of the covert channel, not for exfiltration of sensitive user data, but to check if there are other VMs sharing the same physical machine (defensive detection tool).

The first work that proposed to utilize shared memory for covert communication was introduced by Xiao et al. [9]. Authors described a steganographic method based on memory deduplication. This technique is usually utilized to improve memory efficiency and is typically implemented using a variant of copy-on-write techniques, for which, writing to a shared page would incur a longer access time than to those non-shared. This characteristic feature was utilized to create a local covert channel. If one intentionally increases access delay to some parts of the memory then binary '1' is sent and in the other case binary '0'. The reported steganographic bandwidth was about 80 bps.

As mentioned in the previous section for hard disk sharing Ristenpart et al. [7] introduced a mechanism to determine co-residence of the instances that relies on the shared hard disk contention. To send binary '1', the sender instance reads from random locations on a shared disk volume and to send binary '0' no operation is required. The receiving side infers that '1' was sent from the longer read times and '0' in the other case. The obtained steganographic bandwidth in this case was 0.0005 bps; this means that it takes about 33 minutes to transfer a single bit of secret data.

In this paper we propose CloudSteg which is a significantly improved version of the hard disk contention based method originally proposed by Ristenpart et al. Our experiment revealed that it is possible to achieve a 1000 times higher steganographic bandwidth with good robustness and undetectability.

### III. THREAT MODEL AND ASSUMPTIONS

In this work, similarly as in [7] we consider a public cloud service provider which is trusted as well as this infrastructure. The provider is able to rent computational resources to the users for their own purpose. Users can run one or more copies of VM images and each of the running images is called an *instance*. While such an image starts it is assigned a single physical machine for its lifetime.

We further assume the scenario 3 from Fig. 1 i.e. that there are two cloud instances: one owned by the user which we refer to as *secret data sender (SS)* and the other one which is a *secret data receiver (SR)*. We also assume that the co-residency phase is successfully completed i.e. that it is ensured and verified services, that instances are on the same physical machine (the ways to accomplish this are presented in [7] and [10] ). In such an environment SS and SR utilize the hard disk contention feature specific for cloud computing services to apply the steganographic method that results in a *local covert channel* which allows secret data exchange.

Both users also possess a pre-shared key that allows to define at what data rate secret data bits will be transmitted.

### IV. THE METHOD

To provide successful covert communication CloudSteg utilizes control fields that are required mainly for synchronization purposes. The format of the encapsulated secret message is presented in Fig. 2 and each 8-bit sequence in control fields forms a frame. The secret message to be sent is formed and encapsulated at SS and it starts from two *symbol synchronization* frames (the 16-bits sequence of '1010…'). The next field is a *frame synchronization* sequence (8 bits), which marks the start of user data. After this the user's secret data is inserted ('XYZ…T' in Fig. 2) and then the sequence that marks the end of user data ('00001111') follows. The end sequence is added to be able to carry variable length user data. To ensure that synchronization bits do not mix up with the user data bit stuffing is utilized.

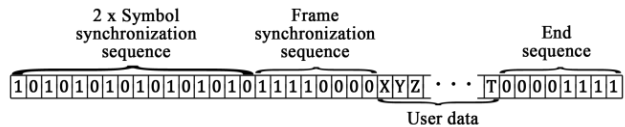

Fig. 2. The format of the encapsulated user data.

Therefore, the basic steps of CloudSteg are as follows. At the SS:
STEP1: Utilize bit stuffing on user secret data.
STEP2: Encapsulate user data with synchronization sequences described above.
STEP3: Choose files that will be used for the hidden data exchange.
STEP4: Encode the resulting secret message into the time change vector (TCV) – for details see next subsection.
STEP5: Send secret message by differentiating files' read times.

At the receiving side SR performs the following steps:
STEP1: Performs measurements of the hard disk access times.
STEP2: Discovers the beginning of the secret message.
STEP3: Decodes the bits of secret message.
STEP4: Performs symbol and frame synchronizations to successfully decapsulate secret message and to extract user data.
STEP5: Removes redundant bits inserted by bit stuffing algorithm.

*A. Secret Data Sender*

As denoted in Section III we assume that two VM instances co-reside on the shared physical machine and in particular use, and share the same hard disk. In such a situation if one of them tries to access its files frantically then it affects the read times of the other instance. This key observation was utilized to enable covert communication in CloudSteg.

When SS wants to send secret data then first he applies bit stuffing on user data and then encapsulates them with synchronization fields (SS steps 1-2).

The next step is to choose the files that are going to be accessed when the covert communication starts. The *n* files ($n \geq 1$) are randomly chosen to be utilized for this purpose (SS step 3). For example, if SS wants to send binary '1' then he accesses the chosen files repeatedly for the predetermined bit time period (known a priori to both communication sides). However if binary '0' is to be sent then SS remains idle for the bit time period. SS tries to access many files instead of a single one to increase the chance of successful decoding of the secret data by SR. It also allows it to cope with sophisticated modern hard disk queuing methods, which may negatively impact the decoding process.

Next, the encapsulated, binary secret message (from step 2) is mapped into time change vector (step 4). TCV reflects the time that the SS should devote to accessing the files or remain idle. For example: if the secret message is [1, 0, 1, 0] and the chosen bit time is 5000 ms then the resulting, corresponding TCV is equal to [5000, 5000, 5000, 5000]. It is also worth noting that every secret message always starts with binary '1' due to the structure of the symbol synchronization sequence (see Fig. 2). Therefore, the TCV can be interpreted as follows: spend 5000 ms accessing the files, then 5000 ms stay idle, etc. Such mapping is especially beneficial if the secret data consist of repeating bit subsequences. Let us consider secret data of [1, 0, 1, 1, 1, 0, 0, 0, 0] and let us assume that bit time is 3000 ms then the resulting TCV is represented as [3000, 3000, 9000, 12000]. Thus, SS first accesses the files for 3000 ms, then stays idle for the same time period. Next, it again accesses the files but for 9000 ms, and finally stays idle for 12000 ms.

It must be emphasised that accessing multiple files requires that each file is accessed in a separate thread. This results in a potential problem when such threads must be stopped when the bit time ends. To avoid extensive files accessing which can negatively affect the secret bits encoding (especially for the large *n*) the following parameter *th* (threshold) is introduced. It specifies how long, before the bit time ends, SS starts to kill files accessing process (kills the threads). For example, if secret data is [1, 1, 1, 1], bit time is 300 ms and *th*=0.9 it means that *th* shortens the time of last bit time to 270 ms to be sure they are all stopped before bit time expires. Therefore, the whole sequence will cause the chosen files to be accessed for 1170 ms. From our observations *th* depends on *n* (for details see Section VB).

*B. Secret Data Receiver*

SR aims at measuring average files' access time in the defined *probing interval* (*pri*). It starts measurements by opening the measurement window which is of a size of *pri* and monitors average hard disk access time in this period.

SR starts inspecting hard disk contention independently of SS. This means that SR is not aware whether SS has already started transmission or not and what secret bit has been sent. SR must decode secret information based on the performed measurements as well as on the predetermined bit time which is known by both covert communication sides.

To be able to successfully detect the beginning of the secret transmission an automatic approach is proposed: SR analyses the incoming secret data in octets and the analysis's phases are illustrated in Fig. 3. These phases are:

*Phase 1:* Detection of the beginning of the secret bit.
*Phase 2:* Determining the binary value for each received bit and global average bit value for all of the received secret bits.
*Phase 3:* Performing symbol synchronization.
*Phase 4:* Performing frame synchronization.

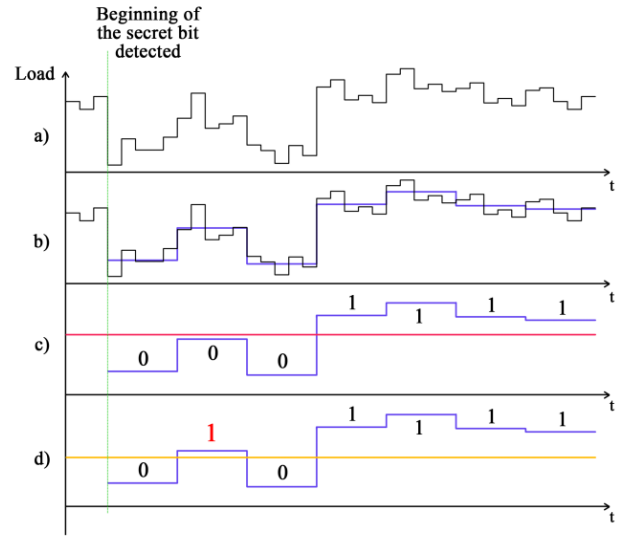

Fig. 3. SR phases: a) detection of the beginning of the secret bit; b) averaging values during single bit time; c) global average bit value; d) corrected global average bit value.

**Phase 1:** Detection of the beginning of the secret bit

Due to the fact that SR does not have any information regarding the start of hidden data transmission it requires a mechanism to automatically detect the start of the secret bit.

To explain the method we propose to address this issue. We define a *subsequence* that is a fixed number of the consecutive measurements' values that has a length of bit time. For example, if the measurement values retrieved by the SR are represented by the set: [A, B, C, D, E, F, G, H] and the length of the bit consist of 3 measurements then exemplary subsequences can be: [B, C, D], [C, D, E], or [E, F, G].

Detection of the beginning of the secret bit is based on determining *subsequences' minimal variance values*. It is based on the observation that subsequences which exactly match the beginning of the transmitted secret bits always have lowest variance values. The example of this concept is illustrated in Fig. 4. We can observe that the calculated subsequences' variance values are always the lowest when they match exactly the beginning of the secret bit.

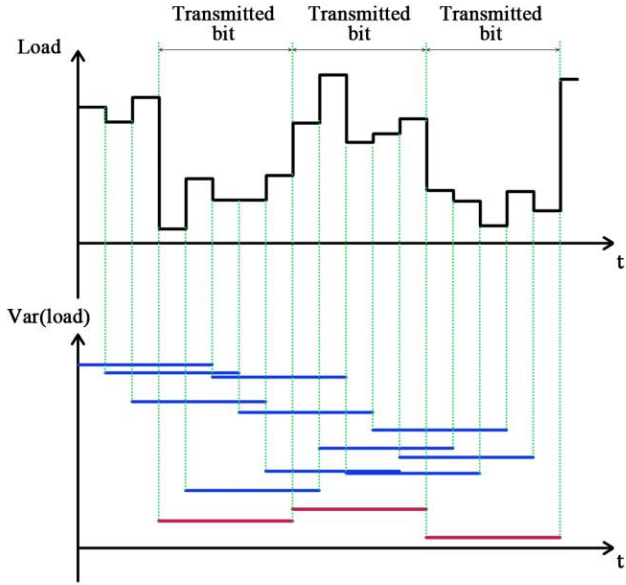

Fig. 4. Detection of the beginning of the secret bit.

To correctly determine the lowest subsequences' variance values, we first define *disjoint* subsequences which we call *upper level subsequences*. For example, if the measurements' results are defined by the set [A, B, C, D, E, F, G, H, I, J, K] and bit time consists of 3 measurements then the upper level subsequences are [A, B, C], [D, E, F], [G, H, I]. We omit subsequences that are shorter than bit time (e.g. [J, K]). Then in the next step we look for the minimum variances' values for each upper level subsequence using so-called *lower level subsequences* i.e. those subsequences that have the same length as upper level subsequence and begin with an element from a certain upper level subsequence. For example, for upper level subsequence [A, B, C] the possible lower level subsequences are: [A, B, C], [B, C, D] and [C, D, E]. Then we calculate the variances for each of these subsequences and choose the one with the lowest value as a bit start (denoted with red lines in Fig. 4).

The last step is to analyse the set of subsequences with the lowest variance values and to choose the one that fits best to represent transmitted bits. It is achieved by utilization of some simple relationships e.g. if there are two disjoint and not neighbouring subsequences, then the time between these subsequences must be equal to bit time period or its multiplicity, etc.

**Phase 2:** Determining the binary values for each received bit

After the previous phase is successfully completed SR is aware of each bit start and stop times. Therefore, based on the measurement values in these periods for each bit time we must determine the resulting binary bit values. The first step it is to calculate average values in each period (Fig. 3b).

Then global average bit value (GAB) is calculated. In an ideal situation GAB is exactly halfway between values of binary '1' and binary '0' (the number of 1s and 0s is the same). During Phase 1 it was assumed that this value is an average for all identified bits (marked with red line in Fig. 3c). In such a case those bits whose average value is higher than GAB are recognized as binary '1'; otherwise they are decoded as binary '0'. However, this is only true if the number of 1s and 0s in the decoded secret data is the same. Otherwise such a situation can lead to secret bit decoding errors. For example, if there are more binary 1s than 0s then the resulting GAB is higher than in an ideal situation where they are equally distributed. To solve this issue the corrected value of GAB must be calculated $GAB_C = GAB + diff$, where *diff* is calculated as follows:

$$diff = 0.5 \,|\, N_1 - N_0 \,| \left( \frac{V_1 - V_0}{N_1 + N_0} \right)$$

where: $N_1$, $N_0$ denotes the number of 1s and 0s in an analysed sequence and $V_1$, $V_0$ are the average measurement values for 1s and 0s.

After the corrected GAB is calculated the binary values for each received secret bit are determined again (Fig. 3d). If the resulting binary values are different as previously calculated, then GAB is corrected again until the values before and after corrections are the same or if the number of 1s and 0s is the same.

**Phase 3:** Performing symbol synchronization

Mechanism for determination of the beginning of each secret bit, based on finding the lowest variances' values introduced earlier (Phase 1) applied alone, is not enough for successful secret data bits extraction. Please consider an example in which the transmitted data bits are [1, 1, 1, 1, 0, 1]. In this case mechanism of determination of beginning of the bits will experience problems as the resulting variances will be similar which may lead to transmission errors. Therefore another mechanism i.e. character synchronization mechanism is needed to synchronize the receiver to the secret data bits transmitted by the sender. To achieve this goal two symbol synchronization frames (see Fig. 2) are added i.e. the sequence of [1, 0, 1, 0, 1, 0, 1, 0, 1, 0, 1, 0, 1, 0, 1, 0]. The synchronization between the sender and receiver is considered successful if the receiver is able to correctly identify at least eight last bits of this sequence.

**Phase 4:** Performing frame synchronization
The frame synchronization mechanism makes it possible to synchronize SS and SR for 8-bits structures and it allows SR to find the beginning and end of the user secret data. Therefore before the user data the sequence of [1, 1, 1, 1, 0, 0, 0, 0] is added and after the user data are transmitted the end of the message sequence is indicated by sequence [0, 0, 0, 0, 1, 1, 1, 1] (compare Fig. 2).

After successful identification of user secret data, SR also needs to remove redundant bits inserted by bit stuffing algorithm at SS.

## V. EXPERIMENTAL EVALUATION

### A. Experimental test-bed

The CloudSteg prototype has been implemented in an OpenStack [12] which is an open source cloud computing environment created by NASA and Rackspace company.

Cloud instances were running on a single OpenStack server therefore (as mentioned before) we assume that the co-residence of the instances determination was successfully established. In this environment two applications were realized: secret data sender and secret data receiver. For determining undetectability of CloudSteg we utilized an additional cloud instance called control probe (for details see Sec. VC). The utilized test-bed is illustrated in Fig. 5.

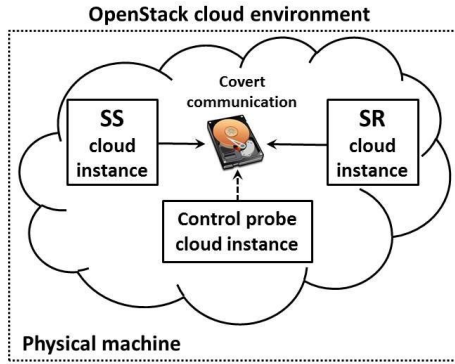

Fig. 5. Experimental test-bed for CloudSteg; SS – secret data sender; SR – secret data receiver.

### B. Performance evaluation measures

CloudSteg performance depends on the following factors:
- Bit time length ($B_T$),
- Probing interval ($pri$),
- $th$ (threshold) parameter,
- The number of accessed files ($n$) utilized for steganographic purposes and $n \epsilon N$.

The relationships between the abovementioned measures are vital to evaluate the performance of the method. For example, the choice of the bit time length affects directly the steganographic bandwidth. Other relationships will be explained below.

**Relationship between *n* and *th***
Both parameters *n* and *th* affect BER (Bit Error Rate) i.e. the correctness of successful receipt of the secret data bits. To obtain the optimal values of *n* and *th* the following experiments were conducted. The measurements were performed in "ideal conditions" (Fig. 5) i.e. that on the utilized server with OpenStack only two cloud instances were active: SS and SR. For these experiments we assumed bit time of 10 seconds to not to introduce too many errors and *pri* = 400 ms. To evaluate BER, 96-bits random sequence was sent three times. The obtained experimental results are illustrated in Fig. 6.

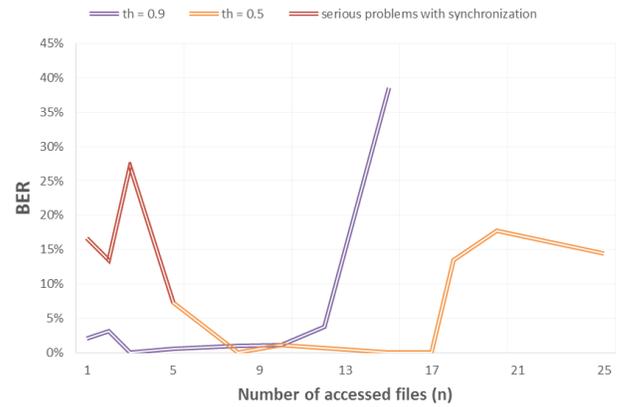

Fig. 6. BER for different values of *n* and *th=0.5 or th=0.9*.

Careful inspection of Fig. 6 leads to the following observations:
- For very small (e.g. for *th*=0.9 it is for *n*<4) and high (e.g. for *th*=0.9 it is for *n*>12) *n* values the BER is the highest, therefore such *n* values should not be utilized for hidden transmission. For *th*=0.5 and *n*=1, 2 or 3 we experienced serious synchronization problems.
- There is a range of *n* values for which BER is almost 0% (e.g. for *th*=0.9 it is for $n \epsilon <4, 12>$).
- For lower *th* values threads are sometimes killed too fast which may also negatively impact BER.

**Determining *probing interval***
From the receiver perspective it is important to correctly assign the probing interval (*pri*) as it influences the secret data extracting process.

The following experiments were conducted to determine *pri* influence on the successful reception of the secret data bits. The secret data of 24 bits were sent under the following conditions: *th*=0.9, *n*=5 and $B_T$=10 000 ms (*pri* is expressed as a percentage of bit time).

Fig. 7 illustrates that for every *pri* value the resulting BER is always close to 0. The only exceptions from this rule are when *pri*:

- *is equal to chosen bit time* which is a result when *pri* starts in the middle of the bit time (it spans through two consecutive secret bit values). Then if values of these two bits are '1' and '0' then the calculated averages are very similar and it is not trivial to successfully extract secret data.
- *is very low* e.g. 0.1-0.2% of bit time.

Based on the obtained results it can be established that *pri* value should not be chosen too low or it should not be equal to bit time.

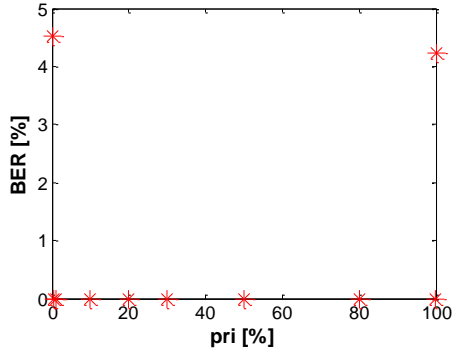

Fig. 7. BER for different values of *pri*.

## C. Experimental results

**Steganographic bandwidth**

Steganographic bandwidth ($S_B$) was measured by transmitting the 96-bits secret bits sequence for every bit time value. The other necessary parameters for experimental evaluation have been assigned as in Table 1.

TABLE I. EXPERIMENTAL RESULTS FOR EVALUATION OF STEGANOGRAPHIC BANDWIDTH

| $B_T$ [s] | 1 | 2 | 3 | 4 | 5 | 8 | 10 |
|---|---|---|---|---|---|---|---|
| $S_B$ [bps] | 1 | 0.5 | 0.33 | 0.25 | 0.2 | 0.125 | 0.1 |
| *pri* [ms] | 40 | 200 | 200 | 200 | 200 | 400 | 400 |
| *n* | 2 | 5 | 5 | 5 | 5 | 5 | 5 |
| *th* | 0.4 | 0.5 | 0.65 | 0.75 | 0.8 | 0.85 | 0.9 |
| *th* [ms] | 600 | 1000 | 1050 | 1000 | 1000 | 1200 | 1000 |

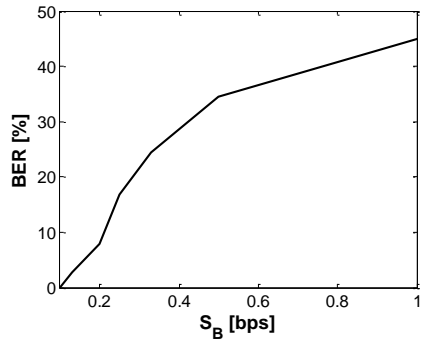

Fig. 8. BER for different steganographic bandwidth values.

Fig. 8 presents experimental results for BER for each value of the steganographic bandwidth evaluated. The proposed method has been able to successfully transmit all secret bits for $S_B$ of 0.1 bps or lower. It means that the chosen $B_T$ should be equal to 10 seconds or lower. This means that the method achieves 1000 times higher steganographic bandwidth than the originally proposed one in [7]. It should also be noted that for higher $S_B$ the resulting BER is higher but e.g. for 0.2 bps it is still below 10%.

**Robustness**

Robustness of the proposed steganographic method shows how the method is able to survive in the presence of unintentional interferences that may occur in the cloud computing environment. For CloudSteg, robustness was evaluated in the two following, separate scenarios: (i) firstly, by adding another cloud instance which focuses solely on verifying its hard disk performance (by utilizing *bonnie++* tool to do the hard disk benchmarking); (ii) secondly, by adding another cloud instance which tries to significantly impact the disk load.

The parameters for the first case experimental evaluation were set as follows: $B_T$=10 s, *n*=5, *th*=0.9, *pri*=400 ms. For every tested scenario the method has achieved BER=0 which means that it is ready to survive such conditions.

The second case was a real stress-test for the proposed method. The cloud instance was trying as often as possible to access the hard disk which significantly impacts the robustness of the steganographic method. Experimental results for this case showed that the average BER was equal to 32.3%. However it must be emphasised that this kind of stress-test is *not* a typical behaviour for cloud instance and should, rather, be treated as intentional.

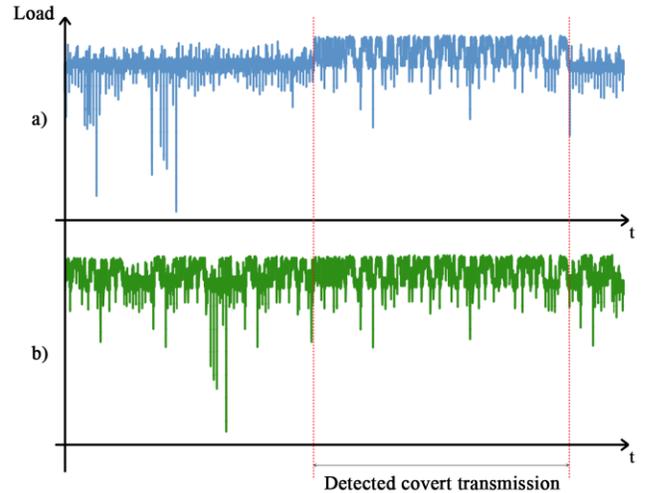

Fig. 9. Undetectability evaluation results in a) ideal scenario; b) potential real-life scenario.

**Undetectability**

For the purpose of undetectability evaluation the control probe which is realized as a separate, third instance is utilized to measure hard disk access times by monitoring its load (see Fig. 5). When the files access time is long then the hard disk load is high and vice versa. Control probe takes the measurements during probing intervals (*pri*) which are constant. Undetectability was evaluated in the following "ideal" scenario: first the control probe was launched and after some random time the covert communication between SS and SR took place.

It is important to note that the control probe as well as SR are not initially synchronized with SS i.e. they are not aware when the transmitted secret bit starts.

The exemplary observed hard disk load diagram is presented in Fig. 9a. By visually inspecting the diagram it can be noticed that there is a time period where the experienced load is higher. Obviously this is a result of the absence of other active cloud instances. If any other cloud instances are active then the resulting hard disk load diagram could be distorted as presented in Fig. 9b thus increasing CloudSteg undetectability.

For cloud provider the potential detection method of CloudSteg could be based on pinpointing such cloud instances which significantly impacts the hard disk load. In such a case it must be emphasised that only SS can be potentially discovered as SR does not affect load at all. Currently some cloud computing platforms allow to monitor parameters of cloud instances e.g. VMware [15] provides triggers that can be activated when the hard disk load is reached by the certain cloud instance as it is defined as a sum of all data read and saved on all cloud instance hard disks in the certain time period. However, even such an approach can be inaccurate in a dynamic cloud computing service due to files' operation of other co-existing cloud instances.

## VI. CONCLUSIONS AND FUTURE WORK

This paper proposes CloudSteg which is an improved steganographic method based on cloud instances hard disk contention originally proposed by Ristenpart et al. in [7]. It allows the covert exchange of secret data between two cloud instances in a cloud computing environment. The method was improved mainly by utilization of different synchronization mechanisms which made it possible to improve resulting steganographic bandwidth 1000 times to about 0.1 bps. The proof-of-concept implementation was realized in an open source OpenStack environment.

Future works include CloudSteg implementation in one of the public, commercially available cloud computing platforms. This could provide a more realistic method's evaluation which is especially important from the undetectability and robustness perspective.

ACKNOWLEDGMENT

This research was partially supported by the Polish National Science Center under grant 2011/01/D/ST7/05054.